\documentclass[journal]{rmaa}
\usepackage{aas_macros}

\usepackage{paralist}

\usepackage{psfrag,color}

\usepackage[latin1]{inputenc}
\usepackage[T1]{fontenc}
\usepackage{rotating}

\title{The variable stars population of the extended young globular
cluster NGC 1851} 

\author{
  A. Arellano Ferro,\altaffilmark{1},
  P\'erez Parra, C.E.\altaffilmark{2,3},
 Yepez, M.A.\altaffilmark{4,5},
  ; Bustos Fierro, I.\altaffilmark{6},
  Prudil, Z.\altaffilmark{7},
  ; Zerpa Guillen, L. J.\altaffilmark{2,3}
}

\altaffiltext{1}{Instituto de Astronom\'ia, Universidad Nacional Aut\'onoma de M\'exico, Ciudad Universitaria, C.P. 04510, M\'exico.}

\altaffiltext{2}{Universidad de Los Andes, Facultad de Ciencias, Dpto. F\'isica, Grupo de Astrof\'isica Te\'orica, M\'erida,
Venezuela.}
\altaffiltext{3}{Fundaci\'on Centro de Investigaciones de Astronom\'ia Francisco, J. Duarte (CIDA), M\'erida, Venezuela.}

\altaffiltext{4}{Instituto Nacional de Astrof\'isica, \'Optica y Electr\'onica (INAOE), Luis Enrique Erro No.1, Tonantzintla, Pue., C.P. 72840, M\'exico}

\altaffiltext{5}{Consejo Nacional de Humanidades, Ciencias y Tecnolog\'ias, Av. Insurgentes Sur 1582, 03940, Ciudad de M\'exico, M\'exico}

\altaffiltext{6}{Observatorio Astron\'omico, Universidad Nacional de C\'ordoba, C\'ordoba C.P. 5000, Argentina.}

\altaffiltext{7}{European Southern Observatory, , Karl-Schwarzschild-Strasse 2, 85748, Garching, Germany.}

\shortauthor{A. Arellano Ferro ET AL.}
\shorttitle{NGC 1851: Physical parameters from variable stars}

\abstract{We report \emph{VI} CCD photometry of the globular cluster cluster NGC 1851. We aim to study the membership of the variable stars detected in the field of the cluster as listed in the Catalogue of Variable stars in Globular Clusters (CVSGC; Clement et al. 2001) and reported by the Gaia mission. We cross match the two sets of variables to produce light curves that lead to the estimation of physical parameters. The resulting colour-magnitude diagram (CMD), free of likely field stars, enables to confirm the position of the variables, their type and evolutionary stage. We provide new estimations of the period using data acquired on a long time-base. The Fourier decomposition of cluster member RR Lyrae light curves lead to a mean metalicity and distance of [Fe/H]$_{\rm ZW}=-1.35\pm0.22$ dex and $11.9\pm 0.6$ kpc. The variability and membership of stars reported by $Gaia$-DR3 as variables in the  field of the cluster is discussed.}

\resumen{}

\addkeyword{Globular clusters: general} 

\addkeyword{Globular clusters: individual (NGC 1851)} 

\addkeyword{Stars: horizontal branch}
\addkeyword{Stars: distances}

\addkeyword{Stars: fundamental parameters}

\addkeyword{Stars: variables: RR Lyrae}

\begin{document}
\maketitle

\section{Introduction}
\label{sec:Intro}

The southern globular cluster NGC 1851 
is a remarkably bright system located in the constellation Columba, and in the outer region of the Milky Way, at a distance of about 12.0 kpc from the Sun ( $\alpha = 5^{\mbox{\scriptsize h}}
14^{\mbox{\scriptsize m}} 06.76^{\mbox{\scriptsize s}}$, $\delta = -40^o 02^{'} 47.6 ^{''}$, J2000; $l = 244.51^o$, $b = -35.03^o$). It is a highly concentrated system \citep{Kuzma2018} characterized by a diffuse halo extending more than 10 times the tidal raidius, although according to \citet{Marino2014}, stars dynamically linked to the cluster are present to at least 2.5 tidal radii. The Galactic orbit of the cluster is very eccentric, e=0.7, and the lack of a tidal tail has triggered the suggestion that the cluster may be a stripped dwarf galaxy nucleus accreted by the Milky Way  \citep{Kuzma2018}. However, a tail that seems to emerge from NGC 1851 was detected by \citet{Carballo2018} which may be interpreted as a tidal remnant of a tentative progenitor dwarf galaxy host of NGC 1851.
Its rather young age, 9.2 Gyrs according to \citet{Koleva2008} or 11.0 Gyrs according to \citet{VandenBerg2013}, seems to support the hypothesis of an extra Galactic origin. From a chemo-dynamical analysis, \citet{Callingham2022} associated NGC~1851 to the ancient major merger event of the Milky Way Gaia-Enceladus-Sausage \citep{Belokurov2018,Helmi2018}.

The horizontal branch (HB) of NGC 1851 posses a moderate population of hot  blue tail stars and a dense red clump nearly twice as populated, hence its HB structural parameter $L=-0.36$ \citep{Arellano2024} is consistent with its Oosterhoff type Oo I and metallicity [Fe/H]=-1.3, following the trend defined by other Galactic clusters of the same Oo-type and similar metallicity.

NGC 1851 harbours a large number of RR Lyrae stars, 48 according the the 2020 edition of the Catalogue of Variable Stars in Globular Clusters (CVSGC) \citep{Clement2001}. It may also contain a 4-5 long period variables near the tip of the red giant branch (RGB). One of our goals in this paper is to employ the variable stars as indicators of the mean metallicity and distance of the parental cluster, hence it is of relevance to ask whether all the variables reported in the CVSGC are cluster members since the large majority of them were discovered before detailed membership analysis was a feasible possibility. In the present paper we propose a membership analyses based on the $Gaia$-DR3 proper motions before we produce a cleaner Colour-Magnitude Diagram (CMD) and hence to study the distribution of the variables and their physical properties. 

\begin{table}[t]
\scriptsize
\begin{center}
\caption{Log of observations of NGC 1851.}
\label{log}
    \begin{tabular}{cccccc} 
    \hline
    Date & N$_V$ & t$_V$(s) & N$_I$ & t$_I$(s) & Mean seeing ($^{\prime \prime}$) \\
    \hline
2018-12-14  & 53 & 100 & 61 & 50 & 3.3  \\
2018-12-15 & 72 & 100 & 80 & 50 & 2.3  \\
2018-12-16 & 62	& 100 & 63 & 50 & 2.7  \\
2019-04-06 & 13 & 100 & 13 & 50 & 3.4 \\
2019-04-07 & 24 & 100 & 29 & 50 & 2.4  \\
2019-04-12 & 23 & 100 & 24 & 50 & 2.5  \\
2019-04-13 & 23 & 100 & 24 & 50 & 2.7 \\
2019-04-28 & 23 & 100 & 24 & 50 & 2.6 \\
    \hline
    Total: & 293 &  & 318  &    \\
    \hline
\end{tabular}
Columns N$_V$ and N$_I$ give the number of images taken with the $V$ and $I$ filters respectively. Columns t$_V$ and t$_I$ provide the exposure time, or range of exposure times. In the last column the prevailing nightly average seeing is listed.
\end{center}
\end{table}

\begin{figure*}
\begin{center}
\includegraphics[width=15.5cm]{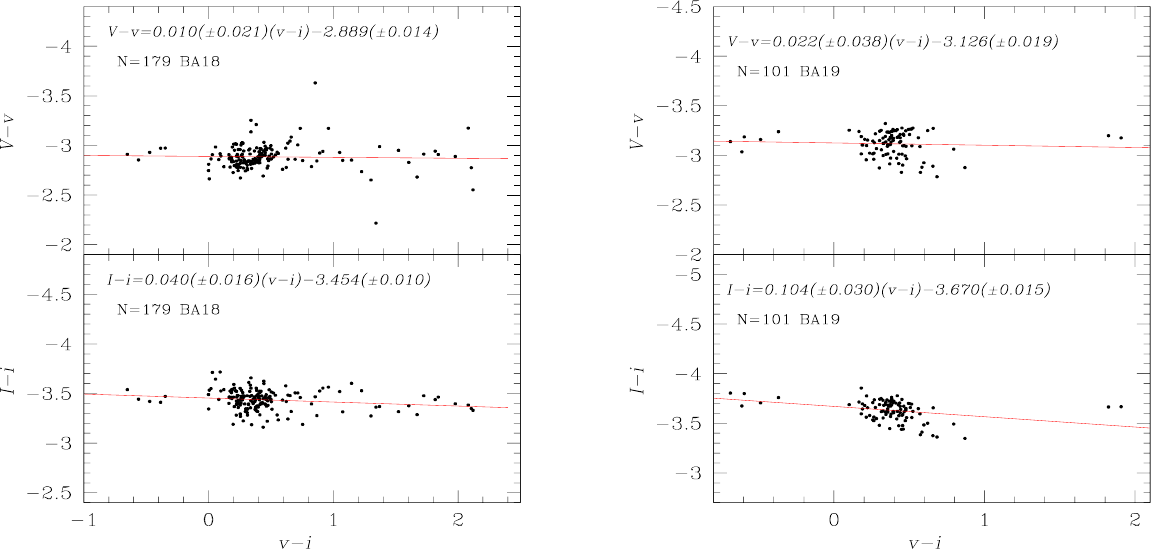}
\caption{Transformation relationship between \emph{VI}   instrumental and standard photometric systems, calculated
for the BA18 and BA19 seasons with 179 and 101 standards respectively, taken from the collection of \citet{Stetson2000}.}
\label{Trans_color}
\end{center}
\end{figure*}

\section{Observations and image reductions}
\label{sec:ObserRed}

\subsection{Observations}

The \emph{VI} CCD images were obtained with the 1.54m telescope of  the  Estaci\'on Astrof\'isica Bosque Alegre del Observatorio de C\'ordoba, Universidad Nacional de C\'ordoba, Argentina (EABA), during 14, 15 and 16 of December 2018 and during five nights between 6-28 in April 2019. We shall refer to these seasons as BA18 and BA19 respectively. 
During the BA18 season we used the camera Alta F16M, equipped with a KAF-16803 chip of 4096$\times$4096 square pixels of 9 microns, binned $2\times2$. This produced a scale of 0.496 arc seconds per pixel and resulted in a field of view of 16.9$\times$16.9 squared arc minutes. During the BA19 season the camera was an Alta U9 with a KAF-6303E CCD detector of 3072$\times$2048 square pixels of 9 microns, also binned 2$\times$2. The scale is 0.496 arc seconds per pixel for a field of view of 12.7$\times$8.5 squared arc minutes.

The log of the observations is given in Table 1, where  the employed exposure times and the nightly seeing conditions are indicated. The goal of these observations is to extract accurate photometry of all point sources in the field of our images, and build their corresponding light curves. For this purpose we employed the difference imaging analysis (DIA) and the DanDIA pipeline \citep{Bramich2008,Bramich2013,Bramich2015}.

\subsection{Locking our photometry to the standard system}
\label{Tranformation}

For the BA18 season we were able to produce light curves for 4623 point sources in the $V$-band and 1578 in the $I$-band. For the BA19 season we measured 2305 light curves in $V$-band and 1949 in $I$-band. These instrumental light curves were transformed to the standard Johnson-Kron-Cousins system defined by \citet{Landolt1992}, by employing local standard stars in the field of NGC 1851 provided in the catalogue of \citet{Stetson2000}\footnote{%
\texttt{https://www.canfar.net/storage/list/STETSON/Standards}}. We identified 179 and 101 standards in BA18 and BA19 respectively for which we have \emph{VI}  photometry. Figure 1 shows the dependence of the standard minus instrumental magnitudes with the instrumental colour $(v-i)$, from which the transformation equations, inscribed in the figure legend, were calculated. These equations were employed to convert all instrumental light curves into the standard system.

Table \ref{tab:vi_phot} displays a small portion of the time-series \emph{VI} photometry obtained in this work. The full table will be made available in electronic form in the Centre de Donn\'es astronomiques de Strasbourg database (CDS).

\begin{table}
\scriptsize
\begin{center}
\caption{Time-series \textit{VI} photometry for the variables stars observed in this work$^*$}
\label{tab:vi_phot}
\centering
\begin{tabular}{cccccc}
\hline
Variable &Filter & HJD & $M_{\mbox{\scriptsize std}}$ &
$m_{\mbox{\scriptsize ins}}$
& $\sigma_{m}$ \\
Star ID  &    & (d) & (mag)     & (mag)   & (mag) \\
\hline
 V1 & $V$& 2458467.53771& 16.000 & 18.889 & 0.014 \\   
 V1 & $V$& 2458467.55788& 16.097 & 18.985 & 0.009 \\
\vdots   &  \vdots  & \vdots & \vdots & \vdots & \vdots  \\
 V1 & $I$ & 2458467.53148 & 15.433 & 18.884 & 0.020\\  
 V1 & $I$ & 2458467.53416 & 15.525 & 18.976 & 0.012  \\ 
\vdots   &  \vdots  & \vdots & \vdots & \vdots & \vdots  \\
 V3 & $V$ & 2458467.53771 & 16.264& 19.155 & 0.019 \\   
 V3 & $V$ & 2458467.55788 & 16.206& 19.097&  0.012 \\
\vdots   &  \vdots  & \vdots & \vdots & \vdots & \vdots  \\
 V3 & $I$ & 2458467.53148 & 15.719&  19.179 & 0.026 \\    
 V3 & $I$ & 2458467.53416 & 15.775&  19.235 & 0.018 \\   
\vdots   &  \vdots  & \vdots & \vdots & \vdots & \vdots  \\
\hline
\end{tabular}
\end{center}
* The standard and
instrumental magnitudes are listed in columns 4 and~5,
respectively, corresponding to the variable stars in column~1. Filter and epoch of
mid-exposure are listed in columns 2 and 3, respectively. The uncertainty on
$\mathrm{m}_\mathrm{ins}$, which also corresponds to the
uncertainty on $\mathrm{M}_\mathrm{std}$, is listed in column~6. A full version of this table is available at the CDS database.

\end{table}

\begin{figure*}[ht]
\begin{center}
\includegraphics[width=17cm]{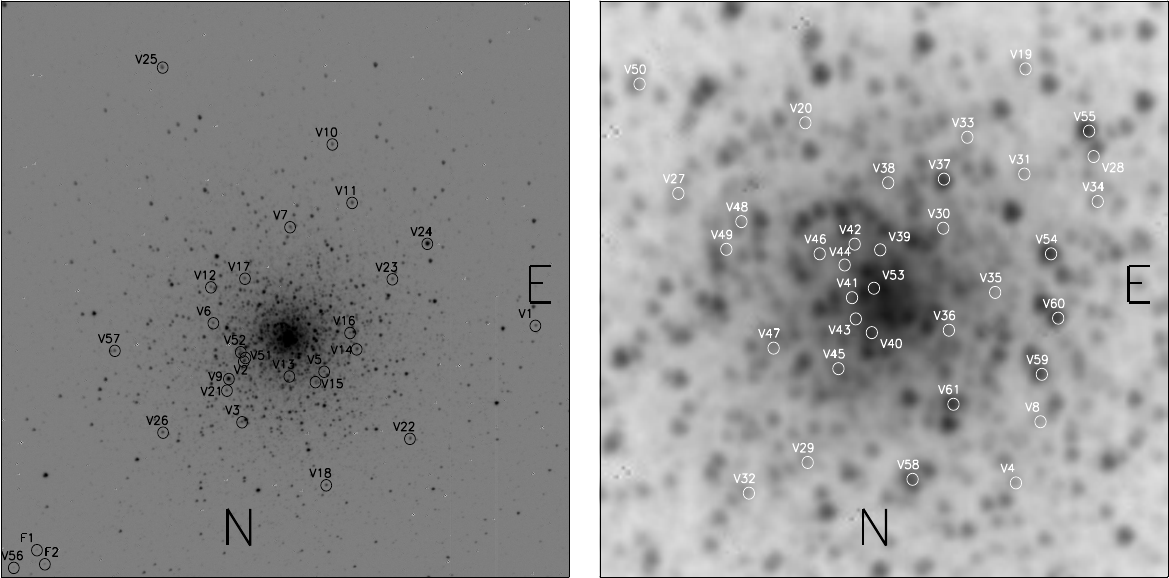}
\caption{Identification chart of variable stars in the field of NGC 1851. The approximate size of the images is $9.7\times9.7$ and $1.6\times1.6$ square arc minutes.}
\label{IdChart}
\end{center}
\end{figure*}

\section{Variable stars in this study}
\label{Vars}

We were able to measure \emph{VI} magnitudes for 1434 stars identified as cluster members in the field of our images. These cluster members will be used to produce the CMD diagram. We could measure  47 of the 55 variables in the cluster listed by \citet{Clement2001} in the CVSGC.
As it can be seen in Table \ref{log}, the observations were carried out under limited seeing conditions and as a result our DIA analysis was unable to retrieve useful data for stars that are very faint, they are near the cluster center or they are blended with a brighter neighbour. We could not measure the RRab stars V30, V36, V39, V40, V41, V43 and V44.  

From and independent exploration we detected clear variability of three faint stars that we temporarily call F1, F2 and F3.

The $Gaia$-DR3 lists 22 variable stars in the field of the cluster that do not match any of the V55 already known in the CVSGC. $Gaia$ photometry is available for 21 of these stars. For the sake of clarity we list these 21 stars in Table \ref{GaiaGroup} along with their $Gaia$ source identification and equatorial coordinates. We arbitrarily numbered these variables with a prefix 'G'.
We explored their light curves from our own photometry and from $Gaia$ data and could confirm with confidence the variability and nature for only five of them.  

All the reported variables  in field NGC 1851 are listed Table \ref{tab:datosgenerales1} along with their variable types, mean magnitudes, amplitudes and ephemerides when this has been possible. Included in the table there are also the stars that we were unable to measure and the non-confirmed $Gaia$ variables  since we are providing their equatorial coordinates and a proper field identification. 

The cluster membership status of all these variables is discussed in the following section.

\begin{table}[t]
\scriptsize
\begin{center}
\caption{Variable stars in the field of NGC 1851 reported in $Gaia$-DR3.}
\label{GaiaGroup}
    \begin{tabular}{cccc} 
    \hline
Var Id &$Gaia$ source  & R.A. & DEC  \\
    \hline
 G1 &  4819198634647024512&  05:14:02.32 & -40:00:01.0\\
G2 &  4819198187968154496 & 05:13:52.49&  -40:01:04.3\\ 
G3 &  4819198089187642112& 05:13:51.14&  -40:02:37.2\\
G4 &  4819197779947169152 & 05:14:02.37&  -40:01:41.9  \\
G5  & 4819197779945787392 &  05:14:00.84 & -40:01:38.4\\  
G6  & 4819197711229437952 &  05:14:05.02 & -40:01:51.9 \\ 
G7  & 4819197676868199168 & 05:14:01.50 & -40:02:37.9 \\ 
G8  & 4819197608151761152 &  05:14:08.52 & -40:01:55.8 \\ 
G9 &  4819197608147820032 &  05:14:11.79 & -40:02:09.2 \\ 
G10 & 4819197505072742912 &  05:14:07.05 & -40:02:18.2 \\
G11 & 4819197505072715264 &  05:14:08.96 & -40:02:35.0 \\ 
G12 & 4819197505072513152 &  05:14:05.82 & -40:02:46.2 \\ 
G13 & 4819197500774037376 &  05:14:09.21& -40:02:44.2  \\
G14 & 4819197500774029312 &  05:14:07.67 & -40:02:30.3 \\ 
G15 & 4819197500774029056 &  05:14:07.64 & -40:02:27.0 \\ 
G16 & 4819197470714194304 &  05:14:02.92&  -40:03:30.8 \\ 
G17 & 4819197436349844992 &  05:14:08.95 & -40:03:40.1 \\ 
G18 & 4819197401991836544 &  05:14:06.29 & -40:03:59.8 \\ 
G19 & 4819197092756042496 &  05:14:01.95 & -40:04:41.6 \\ 
G20 & 4819185822760255616&  05:14:19.10 & -40:02:26.4 \\ 
G21 & 4819185685319251456 &  05:14:13.34 & -40:04:12.9 \\

    \hline
\end{tabular}
\end{center}
\end{table}

We offer a finding chart of all the variables in Fig. \ref{IdChart}. Their light curves are displayed in the Appendix, where we shall  distinguish the data from different seasons. In the Appendix we also discuss individual peculiar or outstanding variables.

\section{Stellar membership analysis}
\label{sec:membership}

In current times, the membership analysis of large numbers of stars in the field of a given globular cluster is on reach thanks to the paramount high quality of proper motions available in the $Gaia$ mission \citep{Gaia2023}. Sieving the likely cluster members and the field stars, enables the production of cleaner CMDs and hence a better perspective of the stellar distributions and evolutionary properties. This is of particular interest for specific groups of variable stars, e.g. RR Lyrae star in the horizontal branch (HB).

The method developed by \citet{Bustos2019} to determine the stellar membership is based on a two step approach: 1) it finds groups of stars with similar characteristics in the four-dimensional space of the gnomonic coordinates ($X_{\rm t}$,$Y_{\rm t}$) and proper motions ($\mu_{\alpha*}$,$\mu_\delta$) employing the  Balanced Iterative
Reducing and Clustering using Hierarchies (BIRCH) clustering algorithm \citep{Zhang1996} and 2) in order to extract likely members that were missed in the first stage, the analysis of the projected distribution of stars with different proper motions around the mean proper motion of the cluster is performed.

\begin{figure*}[ht]
\begin{center}
\includegraphics[width=16.0cm]{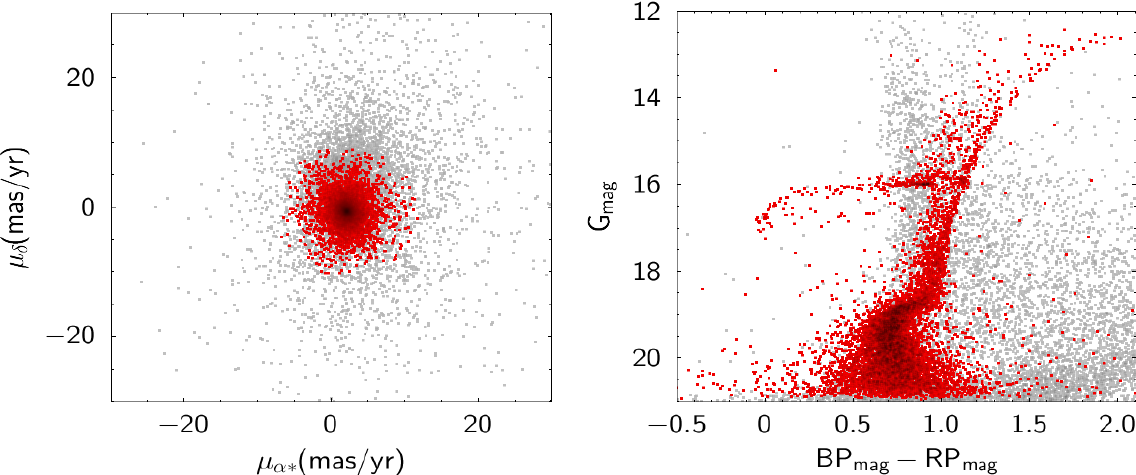}
\caption{$Gaia$-DR3 VPD (left panel) and CMD (right panel) of the cluster NGC 1851. Red and gray points correspond to likely cluster members and field stars respectively, determined as described in section \ref {sec:membership}. A total of 25238 $Gaia$ point sources within 30 arc minutes are displayed, while 11220 were found to be cluster members. }
\label{VPD}
\end{center}
\end{figure*}

Fig. \ref{VPD} shows the corresponding Vector Point Diagram (VPD) and CMD distinguishing the likely cluster members from the field stars. We considered a 30 arc minute radius field from the cluster center, that contains 25238 $Gaia$ point sources out of which 11220 were found to be likely cluster members. 

An independent membership analysis for a large number of globular clusters, based on the proper motions of $Gaia$-DR3, was performed by \citet{Vasiliev2021}. These authors provide membership probabilities for each star in the cluster field. In columns 9 and 10 of Table \ref{tab:datosgenerales1} we list the membership status according the method from \citet{Bustos2019} (B\&C) and the probabilities from \citet{Vasiliev2021} (V\&B). With a few exceptions the match is good. The exceptions that call attention are V8, V25, V37, V42, V44, V45 and V48 and their membership status deserves a few comments. We should note that the coordinates given in the CVSGC are the starting point for a matching with $Gaia$ and that for stars V37, V38, V42, V45, V46 and V48 only X,Y coordinates are listed, hence the matching is sometimes more dubious. In the present work we provide equatorial coordinates for these six stars

We should consider the fact that at the reported coordinates in Table  \ref{tab:datosgenerales1}, we recover in our photometry the light curve of a variable star of the expected type, period and light curve morphology, and that their position in the CMD also becomes a sound membership indicator tool. Therefore, we conclude that V8 is a cluster member, whereas V25, based on its accurate parallax is a field star much closer than the cluster. For RR Lyrae stars V37, V42, V45, V47, V48, V50 and V53 there are good matches with $Gaia$ sources with good proper motions but in all cases the corresponding stars are much brighter than the HB, as can be appreciated in the CMD, hence they are all foreground field RR Lyrae stars. Given that at least one of the two membership identification methods point them as a field stars, we opted for labeling  them as field stars.  Similarly off the HB there are the RR Lyrae V14, V33, and V51, however in  these cases both B\&C and V\&B approaches identified them as very likely cluster members. Since these are positioned towards the central region of the cluster, we cannot ruled out contamination of our photometry by close unresolved neighbours and we opted for considering likely cluster members. We were unable to reliably measure the star V44 in the central region of the cluster, and then its membership status remains unknown (UN).

Regarding the variable stars reported by $Gaia$-DR3, listed in Table \ref{GaiaGroup}, all but G17, G18 and G21 were found to be cluster members by both B\&C and V\&B. All stars marked with 'UN' by B\&C lack proper motion in the $Gaia$ database hence its membership cannot be assessed from that information. In column 11 of Table \ref{tab:datosgenerales1} we list our final membership assessment. 

\subsection{The new variables in the field of NGC 1851}

Of the three newly detected variables in the field of NGC 1851, F1, F2 and F3, only F3 is a likely cluster member, hence we assigned the variable number V56 and we tentativelly classified it as SX Phe star. These three stars are contained in Table \ref{tab:datosgenerales1} and their light curves are displayed in the appendix.

\begin{table*}
    \begin{center}
    \caption{General Data of Variables in the Field of NGC 1851.}
    \scriptsize
    \begin{tabular}{ccccccccccccc}
    \hline
    Variable & Type & $<$\textit{V}$>$ & $<$\textit{I}$>$ & $A_{V}$ & $A_{I}$ & Period & $HJD_{max}$ & B\&C &V\&B &Memb.&RA & Dec.  \\
     &   & (mag) & (mag) & (mag) & (mag) & (days) & (d+2450000) & (M/F) &\%& (m/f/?)&(J2000.0) & (J2000.0) \\
    \hline
\hline
    V1	& RRab	& 16.161  & 15.617 & 1.276 & 0.814	& 0.520583  & 8581.4671	& M1 &0.99 & m& 05:14:28.94 & $-$40:02:56.5 \\
    V2	& CST & ----- & ----- & ----- & ----- & ----- & ----- & M1 &0.99& m&  05:14:02.75 & $-$40:02:24.5\\
    V3	& RRc	& 16.068  & 15.627 & 0.512 & 0.307	& 0.322103  & 8665.7971	& M1 &1.00 & m&  05:14:02.46 & $-$40:01:20.7\\
    V4	& RRab	& 16.197  & 15.617 & 0.770 & 0.447	& 0.585438  & 8602.4723	& M1&0.99  & m&  05:14:08.56 & $-$40:02:17.2\\
    V5		& RRab	& 16.015  & 15.490 & 0.580 & 0.529	& 0.587831  & 7596.6515	& M1 &0.99 & m&  05:14:09.90 & $-$40:02:12.1\\
    V6		& RRab	& 16.094  & 15.537 & 0.916 & 0.579	& 0.606628  & 8468.5608	& M1 &1.00 & m&  05:14:00.02 & $-$40:03:04.4\\
    V7		& RRab	& 16.151  & 15.393 & 1.038 & 0.621	& 0.585186  & 8581.4660	& M1 &0.99 &  m& 05:14:07.04 & $-$40:04:43.4\\
    V8	    & RRab	& 16.154  & 15.566 & 0.939 & 0.519	& 0.510979  & 8469.8129	& M1 &0.16 &  m& 05:14:08.93 & $-$40:02:27.2\\
    V9		& L/SR & ----- & ----- & ----- & ----- & ----- & -----	& M1 &1.00 & m&  05:14:01.34 & $-$40:02:06.0\\
    V10		& RRab	& 16.304  & 15.709 & 0.887 & 0.620	& 0.499528  & 8468.6287	& M1 &1.00 & m&  05:14:10.96 & $-$40:06:09.1\\
    V11		& RRab	& 15.908  & 15.411 & 0.636 & 0.552	& 0.667919  & 8581.5263	& M1 &1.00 & m&  05:14:12.65 & $-$40:05:07.8\\
    V12		& RRab	& 16.223  & 15.684 & 0.957 & 0.580	& 0.575942  & 8586.5057	& M1 &0.99 &  m& 05:13:59.85 & $-$40:03:41.9\\
    V13		& RRc	& 16.170  & 16.172 & 0.623 & 0.558	& 0.282543  & 8580.5020	& M1 &0.61 & ?&  05:14:06.77 & $-$40:02:07.7\\
    V14		& RRab	& 15.413  & 14.697 & 0.376 & 0.204	& 0.594038  & 8469.7261	& M1 &0.89 &  m& 05:14:12.85 & $-$40:02:34.9\\
    V15		& RRab	& 16.140  & 15.353 & 1.184 & 0.782	& 0.541344  & 7002.5754	& M1 &0.99 & m&  05:14:09.11 & $-$40:02:01.4\\
    V16		& RRab	& 16.111  & 15.711 & 1.142 & 0.802	& 0.488699  & 7380.6230	& M1 &1.00 &  m& 05:14:12.28 & $-$40:02:52.1\\
    V17		& RRab	& 16.073  & 15.594 & 0.534 & 0.393	& 0.704841  & 8467.5715	& M1 &0.99 & m&  05:14:02.92 & $-$40:03:50.4\\
    V18		& RRc	& 16.078  & 15.704 & 0.527 & 0.351	& 0.272094  & 8467.5579	& M1 &1.00 & m&  05:14:09.93 & $-$40:00:13.6\\
    V19		& RRc	& 15.852  & 15.392 & 0.409 & 0.281	& 0.405181  & 8602.4853	& M1 &0.99  & m&  05:14:08.79 & $-$40:03:25.3\\
    V19		& RRc	& 15.852  & 15.392 & 0.409 & 0.281	& 0.405181  & 8602.4853	& M1 &0.99 &  m& 05:14:08.79 & $-$40:03:25.3\\
    V20		& RRab	& 16.161  & 15.583 & 0.739 & 0.440	& 0.559460  & 8467.6636	& M1 &0.00 &  m& 05:14:05.58 & $-$40:03:17.0\\
    V21		& RRc	& 16.107 & 15.804 & 0.518 & 0.352	& 0.268520  & 8581.4671	& M1 &0.99 & m&  05:14:01.15 & $-$40:01:53.9\\
    V22		& RRab	& 15.804 & 15.350 & 0.500 & 0.348	& 0.559401  & 8602.4613	& M1 &0.99 &  m& 05:14:17.51 & $-$40:01:00.7\\
    V23		& RRc	& 16.187 & 15.823 & 0.114 & 0.134	& 0.265835  & 8469.7203	& M1 &1.00 & m&  05:14:16.16 & $-$40:03:47.4\\
    V24		& Lb/S	& 13.161 &11.528  & ----- & ----- & ----- & ----- & M1 &1.00 &  m& 05:14:19.35 & $-$40:04:23.9\\
    V25	& EC & 15.706 & 14.823 & 0.461 & 0.445 & 0.173673 & 8587.5103 & M2 &0.00 &  f& 05:13:55.81 & $-$40:07:32.0\\
    V26		& RRc	& 16.149 & 15.732 & 0.482 & 0.305	& 0.328669  & 8469.7001	& M1 &0.99 &  m& 05:13:55.36 & $-$40:01:11.1\\
    V27		& RRab	& 16.063 & 15.532 & 0.978 & 0.661	& 0.523208  & 7440.3531	& M1 &0.99 & m&  05:14:03.72 & $-$40:03:05.7\\
    V28		& RRab	& ----- & ----- & ----- & ----- & ----- & ----- & M1 &0.99 & m&  05:14:09.76 & $-$40:03:10.7\\
    V29		& RRab	& 15.671 & 15.073 & 0.503 & 0.454	& 0.603592  & 8469.7120	& M1 &0.55 &  m& 05:14:05.54 & $-$40:02:21.1\\
    V30		& RRab	& ----- & ----- & ----- & ----- & ----- & ----- & M2 &0.90 &  m& 05:14:07.56 & $-$40:02:59.3\\
    V31		& RRab	& 15.973 & 15.404 & 0.603 & 0.520	& 0.755159  & 7873.3458	& M1 &0.99  & m&  05:14:08.75 & $-$40:03:08.0\\
    V32		& RRab	& 15.909 & 15.073 & 0.488 & 0.284	& 0.659681  & 8602.4747	& M1 &0.99 &  m& 05:14:04.68 & $-$40:02:16.2\\
    V33		& RRc	& 15.680 & 14.818 & 0.588 & 0.126	& 0.341202  & 8468.5790	& M1 &0.99 & m&  05:14:07.93 & $-$40:03:14.2\\
    V34		& RRc	& 15.989 & 15.620 & 0.520 & 0.375	& 0.345033  & 8468.6727	& M1 &0.83 & m&  05:14:09.81 & $-$40:03:03.3\\
    V35		& RRc	& 16.137 & 15.655 & 0.370 & 0.391	& 0.318175  & 7380.6231	& M1 &0.99 & m&  05:14:08.30 & $-$40:02:48.6\\
    V36		& RRab	& ----- & ----- & ----- & ----- & ----- & ----- & UN & U &  ?& 05:14:07.62 & $-$40:02:42.5\\
    V37		& RRc	& 13.217 & 11.769 & 0.066 & 0.046	& 0.351040  & 8469.6989	& UN &1.00 &  f& 05:14:07.56 & $-$40:03:07.4 \\
    V38		& RRab	& 15.885 & 0 & 0.479 & 0.373	& 0.653044  & 8468.7343	& M1 &0.99 &  m& 05:14:06.77 & $-$40:03:07.0 \\
    V39		& RRab	& ----- & ----- & ----- & ----- & ----- & ----- & M1 &0.83 & m&  05:14:06.64 & $-$40:02:55.9\\
    V40		& RRab	& ----- & ----- & ----- & ----- & ----- & ----- & M1 &0.84 & m&  05:14:06.50 & $-$40:02:42.3\\
    V41		& RRab	& ----- & ----- & ----- & ----- & ----- & ----- & UN &U &  ?& 05:14:06.22 & $-$40:02:48.1\\
    V42		& RRc	& 14.648 & 13.638 & 0.360 & 0.163	& 0.309567  & 8580.5193	& M1 &0.20 & f&  05:14:06.27 & $-$40:02:56.9 \\
    V43		& RRab	& ----- & ----- & ----- & ----- & ----- & ----- & M2 &0.99 &  m& 05:14:06.27 & $-$40:02:44.6\\
    V44		& RRab	& ----- & ----- & ----- & ----- & ----- & ----- & UN &0.13 & ?&  05:14:06.12 & $-$40:02:53.5\\
    V45		& RRc	& 14.640 & 13.283 & 0.179 & 0.123	& 0.256363  & 8581.4970	& UN &0.94 &  f& 05:14:06.00 & $-$40:02:36.5 \\
    V46		& RRc	& 14.637 & 14.386 & 0.141 & 0.303	& 0.289664  & 8469.6989	& M2 &0.98 &  f& 05:14:05.76 & $-$40:02:55.4 \\
    V47		& RRc	& 15.629 & 14.703 & 0.369 & 0.148	& 0.280101  & 8602.4680	& M2 &0.59 &  ?& 05:14:05.07 & $-$40:02:40.0\\
    V48		& RRab	& 15.176 & 15.177 & 0.547 & 0.220	& 0.520895  & 8586.5101	& M2 &0.15 & f&  05:14:04.63 & $-$40:03:00.9\\
    V49		& RRc	& 14.174 & 0 & 0.077 & 0.086	& 0.265827  & 8468.5790	& M1 &0.99 & m&  05:14:04.41 & $-$40:02:56.5 \\
    V50		& RRc	& 15.462 & 14.787 & 0.231 & 0.135	& 0.325064  & 8581.5275	& M1 &0.00 &  f& 05:14:03.18 & $-$40:03:23.8\\
    V51		& RRab	& 14.805 & 13.899 & 0.245 & 0.214	& 0.509389  & 8580.5303	& M1 &1.00 & m/f?&  05:14:02.75 & $-$40:02:24.5\\
    V52	   & RRab	& 16.120 & 15.327 & 0.256 & 0.115	& 0.648831 & 8468.6052	& M1 &0.99 &  m& 05:14:02.44 & $-$40:02:33.8\\
    V53		& RRc	& 12.617 & 11.260 & 0.260 & 0.257	& 0.325140  & 8581.4789	& UN &0.99 & f&  05:14:06.54 & $-$40:02:49.6\\
    V54		& L	& ----- & ----- & ----- & ----- & ----- & ----- & M1 &0.99 &  m& 05:14:09.12 & $-$40:02:54.8\\
    V55		& L & ----- & ----- & ----- & ----- & ----- & ----- & M1 &1.00 &  m& 05:14:09.70 & $-$40:03:14.9\\
\hline
    \hline
    \end{tabular}
    \label{tab:datosgenerales1}
    \end{center}
\end{table*}

\begin{table*}
\addtocounter{table}{-1}
    \caption{Continue}
    \scriptsize
    \begin{center}
        \begin{tabular}{ccccccccccccc}
            \hline

    Variable & Type & $<$\textit{V}$>$ & $<$\textit{I}$>$ & $A_{V}$ & $A_{I}$ & Period & $HJD_{max}$ & B\&C & V\&B &Memb.&RA & Dec.\\
     &   & (mag) & (mag) & (mag) & (mag) & (days) & (d+2450000) & (M/F) & \%&(m/f/?)&(J2000.0) & (J2000.0) \\
   \hline
   \hline
\multicolumn{11}{c}{New Stars Identified in the NGC1851 Field.}\\
\hline
V56		& SX Phe? & 18.381 & 17.842 & 0.273 & 0.28 & 0.250666 & 8468.5802 & M1  &1.00& m& 05:13:41.80 & $-$39:58:52.3\\
F1		& RRc & 18.419 & 17.282& 0.236& 0.455  & 0.337433 & 8467.5760 & FS  &0.00  &f& 05:13:43.90 & $-$39:59:10.4\\
F2		& RRc & 17.843 & 16.100 & 0.236 & 0.296 & 0.257364 & 8467.5660 & FS  &0.00&f& 05:13:44.59 & $-$39:58:55.6\\
    
    \hline
\multicolumn{12}{c}{Variables in $Gaia$-DR3 confirmed in the present work}\\
\hline
V57(G3)		& RRab & 16.106 & 15.509 & 0.141 & 0.122 & 0.714154 & 8468.6053 & M1  &1.00& m& 05:13:51.14 & $-$40:02:37.2\\

V58(G10)		& RRab & 14.352 & 13.211 & 0.197 & 0.126 & 0.503017& 7063.5570 & M1  &1.00&  m&05:14:07.05 & $-$40:02:18.2\\

V59(G11) & L & 13.55 & 12.03 & 0.40 & 0.21 & ----- & ----- & M1 &  1.00&  m&05:14:08.96 & $-$40:02:35.0\\
V60(G13) & L  & 13.42 & 11.63 & 0.37 & 0.20 & ----- & ----- & M1 &1.00 &  m&05:14:09.21 & $-$40:02:44.2\\
V61(G14) & L & 13.27 & 11.62 & 0.43 & 0.19 & ----- & ----- & M1 &  1.00&  m& 05:14:07.67& $-$40:02:30.3\\

    \hline
\multicolumn{12}{c}{Variables in $Gaia$-DR3 not confirmed in the present work}\\
\hline
 & &  & & & & & & & & & & \\
 & & $\overline{V}$ & $\overline{I}$& & & & & & & & & \\
\hline
 
G1 	& ----- & 19.129& 18.268  & ----- & ----- & ----- & ----- &M1 & 1.00& m&    05:14:02.32 & $-$40:00:01.0 \\
G2&----- &  18.952& 18.593& ----- &----- &-----  & ----- & M1  &1.00& m& 05:13:52.49 & $-$40:01:04.3\\
G4		& ----- & 17.452  & 16.467 & ----- & ----- & ----- & ----- & M1 &1.00 &  m&05:14:02.37 & $-$40:01:41.9\\
G5 	& ----- &$\it {19.461}$ & $\it {18.610}$ & ----- & ----- & ----- & ----- &M1&1.00 &  m&05:14:00.84    & $-$40:01:38.4 \\
G6		&-----  & $\it {19.406}$ &  $\it {18.756}$& ----- & ----- & ----- & ----- & M1 &1.00 & m& 05:14:05.02 & $-$40:01:51.9\\
G7 	&-----  & $\it {19.299}$ & $\it {18.644}$ & ----- & ----- & ----- & ----- &M1 &1.00 & m&05:14:01.50   & $-$40:02:37.9  \\
G8 	&-----  & 17.027 & 16.138 & ----- & ----- & ----- & ----- &M1&1.00 &  m&05:14:08.52  & $-$ 40:01:55.8 \\
G9 	& ----- & 18.585& 18.118& ----- & ----- & ----- & ----- &M2 &0.00 & f&05:14:11.79    & $-$40:02:09.2 \\

G12	&----- & 13.442 & 11.815 & ----- & ----- & ----- & -----& M1 &1.00 & m& 05:14:05.82 & $-$40:02:46.2\\

G15	& -----& 13.230 & 11.513 & ----- & ----- & ----& ----& M1 &1.00 & m& 05:14:07.64 & $-$40:02:27.0\\

G16 &-----  &$\it {19.629}$&$\it {18.943}$ & ----- & ----- & ----- & ----- & M2 &0.99&  m&05:14:02.92 & $-$40:03:30.8\\

G17&-----  & 16.672 & 15.754  & ----- & ----- & ----- & ----- & UN &0.89&  ?&05:14:08.95 & $-$40:03:40.1\\
G18 	&-----  & $\it {19.601}$ & $\it {18.926}$ & ----- & ----- & ----- & ----- &M2&0.01 & ?&05:14:06.29   & $-$40:03:59.8 \\
G19 	&-----  & 15.472&14.510 & ----- & ----- & ----- & ----- &M1&1.00 & m&05:14:01.95  & $-$40:04:41.6 \\
G20	&-----  & $\it {20.219}$ &$\it {19.732}$& ----- & ----- & ----- & ----- &M1&1.00 &  m&05:14:19.10  & $-$40:02:26.4 \\

G21		&-----& 17.363 &16.544 & ----- & ----- &----- & ----- & M2 &0.00 &  ?&05:14:13.34 & $-$40:04:12.9\\

 \hline
    \end{tabular}
    \center{$^1$      Columns 3 and 4 contain intensity weighted means, except for the LPV stars and the G-group in the bottom section, which are magnitude weighted means. Numbers in italic are averages exclusively from $Gaia$ data transformed into \emph{VI}. Columns 5 and 6 are light curves amplitudes. Column 9 indicates the membership status found in this work from the method of  \citet{Bustos2019} (M1 or M2 for likey members, UN for unknown and FS for field stars). Column 10 contains the membership probability assigned by \citet{Vasiliev2021}. }\\
    \end{center}
\end{table*}

\section{The Oosterhoff type of NGC 1851}

\begin{figure}[ht]
\begin{center}
\includegraphics[width=6.0cm]{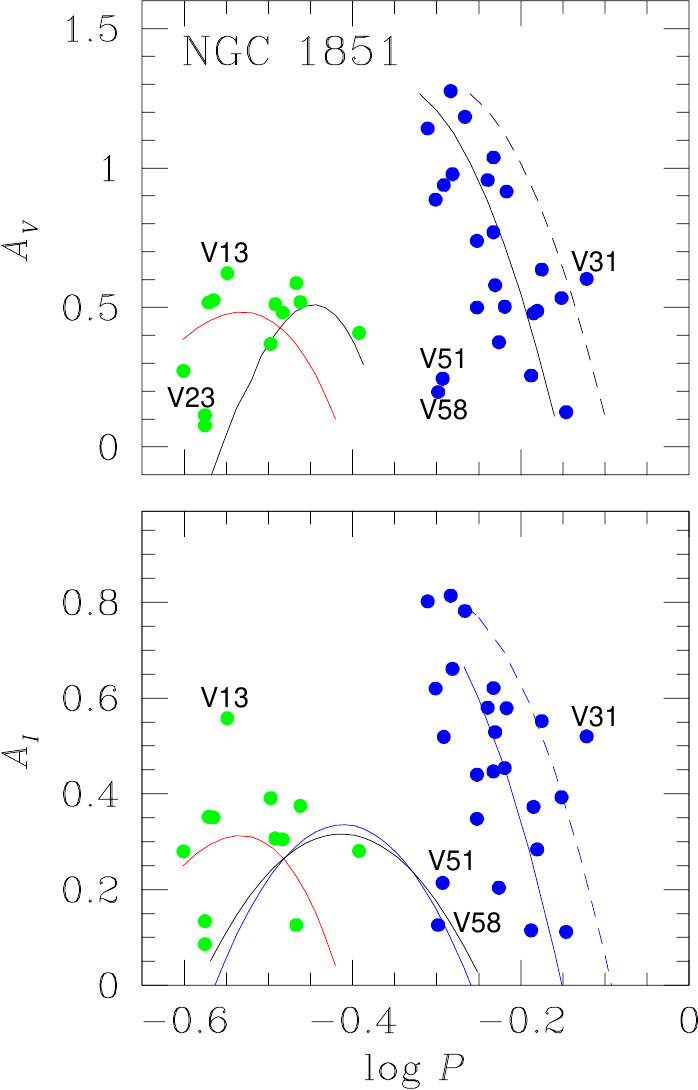}
\caption{Figure 4. Period-Amplitude diagram for RR Lyrae stars in NGC 1851. Blue and green circles represent RRab and RRc stars respectively.}
\label{Bailey}
\end{center}
\end{figure}

The period averages for the RRab and RRc  stars in NGC 1851 are <Pab> = $0.57\pm 0.06$ days and <Pc> = $0.31\pm 0.04$ days. These numbers point to an Oostehoff type Oo I for this cluster. In the log P - amplitude plane, or Bailey's diagram, of Fig. \ref{Bailey} we plot amplitudes and periods for all RR Lyrae stars measured in the this work (Table \ref{tab:datosgenerales1}). The distribution of star on this plane clearly favours the unevolved sequences and the Oo I type of this cluster, in consistency with the averages periods and the metallicity of the cluster of about [Fe/H]$_{\rm UV} = -1.25$, (see \S \ref{sec:Four}).   

\section{On the cluster reddening}

 Given its Galactic location, the reddening of NGC 1851 is relatively low compared to other globular clusters closer to the galactic plane, where the density of dust and gas is higher. Independent estimates of the cluster reddenig consistently report low values of $E(B-V)$; 
  for example, \citet{Harris1996} and \citet{WALKER1998} give a value of $E(B-V)$ = 0.02, the latter stresses that there are no compelling evidences for values too different from this estimation. The calibrations of 
\citep{Schlegel1998},
\citep{Schlafly2011} give a values of 0.037 and 0.032 respectively. 
We have not attempted a reddening determination from the colour $(V-I)$ being constant between phases 0.5-0.8 for RRab stars \citep{Sturch1966} and the subsequent calibration of \citet{Guldenschuh2005}, since our $I$ light curves are scanty and phase gaps are present. Hence we adopted the value of $E(B-V)$ of 0.032 $\pm$ 0.002 from the calibration by 
\citep{Schlafly2011}.

\section{Physical Parameters of the RR Lyrae Stars from the light curves Fourier decomposition}
\label{sec:Four}

The Fourier decomposition of the light curves of RR Lyrae stars, both RRab and RRc, is a well established approach towards the determination of some physical parameter, mainly the metallicity [Fe/H], the luminosity (or absolute magnitude $M_{\rm V}$ and hence the distance), as well as the mass and mean stellar radius.
The Fourier decomposition technique as well as the semi-empirical calibrations and their zero points leading to the physical parameters, have been presented and discussed in detail in the papers by
\citet{Arellano2010}. A summary of the results for 40 clusters calculated homogeneously for over more than a decade can be found in \citet{Arellano2024}. The interested reader is referred to those works for the the involved details.

In the present paper we have limited the calculation of physical parameters to those stars that have proven to be likely cluster members, according to the discussion offered in \S \ref{sec:membership} and summarized in Table \ref{tab:datosgenerales1}.

\subsection{Physical parameters of RR Lyrae stars}

In Table \ref{tab:parfisRR} there are given the values of [Fe/H] in the scales of \citet{Zinn1984} and in the spectroscopic scale of \citet{Carretta2009} for the member RRab and RRc stars. Also listed are the individual values of log~$T_{\rm eff}$, log$(L/{L_{\odot}})$, $M/{ M_{\odot}}$, $R/{ R_{\odot}}$ and distance. 
All the reported mean values in this table have been weighted by the inner uncertainties, which are given between parethesis as described in the notes at the bottom of the table.
The average [Fe/H] and distances are considered good representations of the metallicity and distance of the parental cluster. We find [Fe/H]$_{\rm ZW}=-1.35\pm0.22$, or 
in the spectroscopic scale of \citet{Carretta2009} [Fe/H]$_{\rm UV}=-1.16\pm0.25$ and a distance to the cluster of $11.9\pm 0.6$ kpc.

\begin{table*}[htp]
\footnotesize
\begin{center}
      \caption{Physical parameters from the member RR Lyrae Fourier light curve decomposition.}
    \begin{tabular}{ccccccccc} 
    \hline
    ID &[Fe/H]$_{\rm ZW}$ & [Fe/H]$_{\rm UVES}$  & $M_V$ & log~$T_{\rm eff}$  &log$(L/{L_{\odot}})$ &$M/{ M_{\odot}}$ & $D(kpc)$ &$R/{ R_{\odot}}$ \\
    \hline
    \multicolumn{9}{c}{RRab}\\
    \hline
    V1   &-1.49(3)$^1$ & -1.40(3)   &  0.57(1) & 3.82(1) & 1.68(1) &0.74(7) & 12.55(3) & 5.42(1) \\ V6  &-1.25(5) & -1.14(4)   & 0.54(1) & 3.81(1) & 1.69(1) & 0.67(8) & 12.35(3) & 5.69(1) \\V7   &-1.40 (4) & -1.29(4)   & 0.53(1) & 3.81(1) & 1.69(1) & 0.68(7) & 12.73(3) & 5.63(1) \\V11 &-1.32(34)$^2$ &   -1.20(32)$^2$   & 0.52(1) & 3.80(1) & 1.70(1) & 0.62(11) & 11.44(1) & 5.85(1) \\  
    V12  &-1.47(6)   & -1.38(7)  & 0.55(1) & 3.81(1) & 1.69(1) & 0.70(11) & 13.00(4) & 5.61(2) \\ 
    V15  &-1.43(9)  & -1.33(9)   & 0.61(1) & 3.82(2) & 1.66(1) & 0.68(12) & 12.22(6) & 5.35(2) \\   
    V16  &-1.05(6) & -0.94(5)   & 0.60(1) & 3.82(1) & 1.66(1) & 0.68(9) & 12.42(4) & 5.07(2) \\     V38   &-1.41(12)  & -1.31(12)  & 0.55(1) & 3.84(3) &1.69 (1) & 0.40(15) & 11.34(3) & 4.87(1) \\  
    
    \hline
    Mean &-1.38 & -1.25 & 0.54 & 3.81 & 1.68& 0.67 & 11.92 & 5.52\\
        $\sigma$ & $\pm 0.15$  &$\pm 0.15$ & $\pm 0.03$ & $\pm 0.01$ & $\pm 0.02$  & $\pm 0.11$ &  $\pm 0.68$ & $\pm 0.33$ \\
    \hline
    \multicolumn{8}{c}{RRc}\\
    \hline
    V3 &-1.56(22)  & -1.49(24)   & 0.57(1) & 3.83 (1) & 1.67(1)  & 0.78(1)  & 12.06(3) &4.96 (1) \\
     V9 &-1.81(29)  & -1.82(38)   &0.51 (1) & 3.83 (1) & 1.70(1)  & 0.63(1)  & 11.21(5) & 5.23(3) \\ 
   V21 &-0.96(16)  & -0.87(11)   &0.63 (1) & 3.88 (1) & 1.65(1)  & 0.59(1)  & 11.95(6) & 4.00(1) \\ 
 V23 &-1.45(47)  & -1.35(48)   &0.75(2) & 3.86 (1) & 1.60(1)  & 0.63(2)  & 11.70(12) & 4.09(4) \\ 
    V26 &-1.36(21)  & -1.25(21)   &0.54(1) & 3.84 (1) & 1.68(1)  & 0.71(1)  & 12.67(6) & 4.85(2) \\ 
    V34 &-1.07(79)  & -0.96(61)   &0.49 (1) & 3.86(1) & 1.70(1)  & 0.59(2)  & 12.04 (8) & 4.64(3) \\

    \hline
    Mean & -1.29 &	-1.07 & 0.59 & 3.85 &  1.65 & 0.64& 11.96  & 4.42\\
    $\sigma$ & $\pm 0.31$  &$\pm 0.35$ & $\pm 0.10$ & $\pm 0.02$ & $\pm 0.04$ & $\pm 0.08$ &  $\pm 0.48$ & $\pm 0.49$ \\
    
    \hline
    \end{tabular}
    \center{$^1$ Numbers in parentheses indicate the internal uncertainty expressed to the last digit; e.g. -1.49(3) is equivalent to -1.49$\pm$0.03}.
    
    $^2$ Value not included in the mean.
    \label{tab:parfisRR}
    \end{center}
\end{table*}

\section{The Color Magnitude Diagram}
\begin{figure*}[t]
\begin{center}
\includegraphics[width=16.0cm]{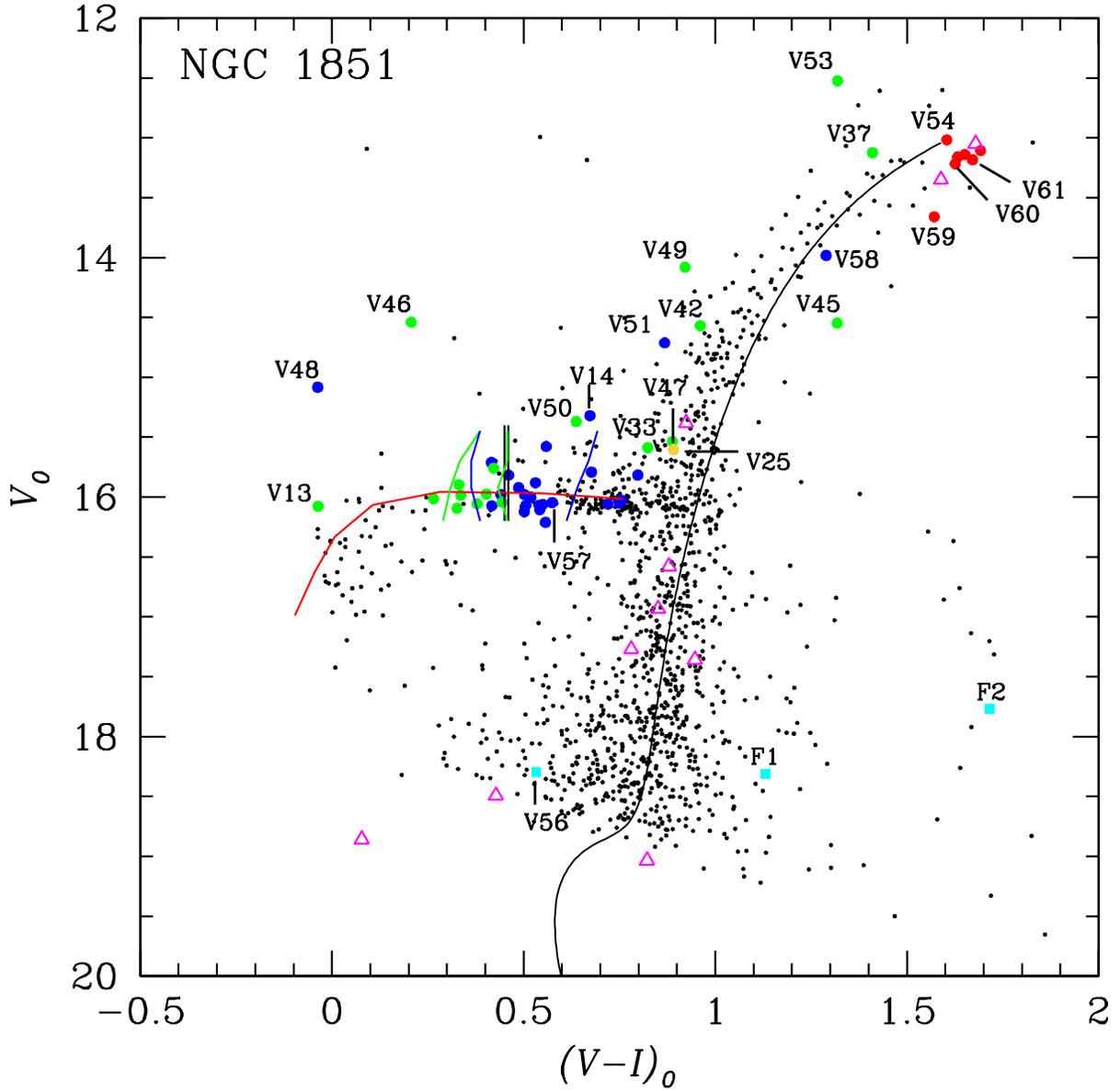}
\caption{Color-Magnitude Diagram (CMD) of NGC 1851. Variables stars in the field of the cluster are plotted with colour symbols according to the following code:solid blue and green circles represent RRab and RRc star respectively; red cicles are for SR/L variables near the tip of the RGB. The star V25, classified as an eclipsing binary, is shown with a yellow circle. Tuquoise colour is use for three newly identified variables, and purple open triangles for variables reported in $Gaia$ that were not identified in our photometry or were not confirmed as variables. The Red ZAHB was constructed by \citet{Yepez2022} using the models built from with the Eggleton \citep{Pols1997,Pols1998,KPS1997}. The green and blue vertical nearly vertical lines on the HB are the theoretical first overtone and fundamental mode instability strips respectively \citep{Bono1994}.en
The isochrone is from VandenBerg et al. (2014) for
[Fe/H]=-1.35 and an age of 12.0 Gyrs. The vertical black lines at the ZAHB mark the empirical red edge of the first overtone instability strip \citep{Arellano2015, Arellano2016}.}
\label{CMD}
\end{center}
\end{figure*}

The observed  CMD  of NGC~1851 built from our \emph{VI} photometry only with likely cluster member stars, was dereddened assuming $E(B-V)=0.03$ mag. The resulting intrinsic CMD is displayed in Fig. \ref{CMD}. All variable stars are plotted with the colours and symbols code in the caption. This diagram helps confirming the non membership of many stars as discussed in previous sections since their positions are at odds with their variable type in many cases. We remind the reader that our final membership assessment is given column 11 of Table \ref{tab:datosgenerales1}.

We call attention at this point to the distribution of RRab and RRc stars on the HB. Considering exclusively the stars that are clear cluster members, we see that some RRab are seating in the bimodal region of the instability strip, i.e the intersection  of the fundamental and first overtone instability strips. The theoretical bounds of these strips are indicated by the green and blue borders calculated by \citet{Bono1994}. The empirical border of the first overtone red edge (FORE) is indicated by the two vertical black lines in te HB \citep{Arellano2015,Arellano2016} and matches well with the theoretical FORE.
RRab stars crossing to the blue of the FORE is a characteristic of some Oo I type clusters, like NGC~1851, but it does not happen in Oo II clusters where the RRab remain to the red of the FORE, i.e. off the bimodal region (see \citet{Yepez2022} and references there in for a discussion). This characteristic of Oo II clusters is probably a consequence of the more advance stage in their evolution to the red, towards the AGB.

Of the 21 stars in the field of NGC 1851 marked in $Gaia$-DR3 as variables, we found a counterpart measured in our photometry for 15 of them.  The others,  were either too faint or unresolved given the spatial resolution of our images.
For a proper comparison we transformed the $Gaia$ photometric data into \emph{VI} using the transformation equations of \citep{Riello2021}. We could confirm the variability and variable type in 5 of them; V57-V61 in Table \ref{tab:datosgenerales1}. The remaining 10 are plotted in the DCM with purple open triangles and are distributed all across the diagram. The $Gaia$ cadence is not designed for the identification of some variables, therefore the authenticity of these variables will have to be confirmed with a proper time-series observations on images of higher resolution than ours.  

The theoretical ZAHB shown in the figure with a red continuous locus, was calculated  by \citet{Yepez2022} using the models built with the Eggleton code \citep{Pols1997,Pols1998,KPS1997} for a metallicity of z=0.001, a core mass of 0.5 $M/{ M_{\odot}}$, and  a range of total masses of 0.59-0.68 $M/{ M_{\odot}}$. The isochrone is from VandenBerg et
al. (2014) for [Fe/H]=-1.35 and an age of 12.0 Gyrs.

 All the above loci have been drifted to a distance of 11.95 kpc and represent well the observed ditribution of the cluster member stars.

\section{Conclusions}

The presence of variable field stars projected against  a Galactic globular cluster is very common, and while such contamination by field stars in the Galactic bulge globular clusters can be remarkably high \citep[e.g. see the case of NGC 6558][]{Arellano2024b} due to the richness of the bulge of field variable stars, particularly of RR Lyrae, it can also be noticeable in more isolated globular clusters in the outer regions of the Milky Way. Such is the case of NGC 1851 as we have demonstrated in the present work. Ad hoc membership analysis based on the proper motions and parallaxes available in $Gaia$-DR3, complemented with mean magnitudes and colours in the $V -(V-I)$ CMD, has shown that of the 55 variables originally listed in the CVSGC, 8 have been found to be clearly field stars and, for 6 more the membership cannot be solidly assessed due to the lack of proper motion or to blending with bright neighbours, particularly in the central regions of the cluster.

Three variables not detected before were identified but only one showed to be a cluster member. We named it V56 and classified it tentatively as an SX Phe star. Among the 21 variables reported by $Gaia$ not included in the CVSGC, we confirmed the variability of two RRab and three long term L variables. Since they turned out to be cluster members we asign them variable names V57-V61.

Identifying variable cluster members is rewarding since they can be used, with more confidence, as indicators of average physical quantities representative of the parental cluster. Here we have estimated the mean metallicity and distance of NGC 1851 via the Fourier decomposition of RR Lyrae light curves, to find [Fe/H]$_{\rm ZW}=-1.35\pm0.22$ dex and $11.9\pm 0.6$ kpc.

A few comments on the position of NGC 1851 relative to the Oosterhoff gap are in order, since the cluster has been associated with an CMa dwarf galaxy \citep{MArtin2004}. We noted before that the average period of the member RRab stars is <Pab> = $0.57\pm 0.06$ d, which with the metallicity [Fe/H]$_{\rm ZW}=-1.35$ places the cluster among the Oo I clusters and slightly off the Oosterhoff gap marked by \citet[][see his figure 5]{Catelan2009}. On the other hand, let us consider the structural, or Horizontal Branch type parameter, defined as $HBt=(B-R)/(B+V+R)$ \citep{Lee1994}, where $B$ and $R$ are the number of star to the blue and to the red of the instability strip respectively, and $V$ represents the number of RR Lyrae in the instability strip \citep{Lee1994,Demarque2000}. In the [Fe/H]-$HBt$ plane \citet[][his figure 7]{Catelan2009} identified a region void of Galactic globular clusters, but populated otherwise by cluster associated with neighbouring galaxies, and termed this region as "forbidden" or the "Oosterhoff gap?". 
We should recall here that the Oo I clusters NGC 1851 and NGC 2808, as well as the Oo II  clusters NGC 2298 and NGC 1904, have been suggested by \citet{MArtin2004} to be associated to the Canis Major dwarf galaxy accreted by the Milky Way. More recently, \citet{Callingham2022} have associated the first three to the Gaia-Enceladus-Sausage merger event and to the Helmi merger \citep{Helmi2018} for the case of NGC~1904.

NGC 1851 and NGC 2808 have well developed HB blue tails but prominent red clumps, hence their  $HBt$ values are very red, i.e. negative, whereas NGC 1904 and NGC 2298 have massive blue tails but lack a red clump and hence their $HBt$ values are very blue, hence large and positive. 

Considering the updated version of the [Fe/H]-$HBt$ plane \citep[][see his figure 11]{Yepez2022}, and plotting these four clusters with the coordinates ([Fe/H],$HBt$); NGC~1851 (-1.35,-0.36 this work), NGC~2808 (-1.15, -0.49 \citet{Catelan2009}), NGC~1904 (-1.68,+0.74 \citet{Arellano2024}) and NGC~2298 (-1.96,+0.96 \citet{Torelli2019}), it is evident that none of these four clusters occupy the Oosterhoff gap.

We are faced with two possible conclusions; these clusters are not associated to external galaxies mergers of the MW beyond the spatial coincidence, or else the globular  clusters of extragalactic origin can occupy regions in the [Fe/H]-$HBt$ or [Fe/H]-$<P_{\rm ab}>$ planes other than the Oosterhoff gap defined by \citet{Catelan2009} as in fact some are seen in his figures 5 and 7. This reinforces that the Oosterhoff gap retains its meaning only in Galactic terms. 
Hence, we do not find a compelling evidence, from this argumentation, for an association of NGC 1851 (and perhaps neither of NGC~2808, NGC~1904 and NGC~2298) to the large accretion events that seem to have sculpt the Galactic halo.

\vskip 1.0cm

\section{ACKNOWLEDGMENTS}

AAF is grateful to the European Souther Observatory (Garching), for warm hospitality during the writing of this work. The permanent support from the IA-UNAM librarian, Beatriz Ju\'arez Santamar\'ia, with the bibliographical material needed for this work is fully acknowledged. AAF also thankfully acknowledges the sabbatical support granted by the
program PASPA of the DGAPA-UNAM. We have been benefited from the support of DGAPA-UNAM through projects IG100620 and IN103024.

\bibliography{NGC1851}

\begin{thebibliography}
\expandafter\ifx\csname natexlab\endcsname\relax\def\natexlab#1{#1}\fi
\expandafter\ifx\csname href\endcsname\relax
  \def\href#1#2{}\fi
\expandafter\ifx\csname urllinklabel\endcsname\relax
  \def\urllinklabel{[LINK]}\fi
\expandafter\ifx\csname adsurllinklabel\endcsname\relax
  \def\adsurllinklabel{[ADS]}\fi

\bibitem[{{Arellano Ferro}(2024)}]{Arellano2024}
{Arellano Ferro}, A. 2024, IAU Symposium, 376, 222


\bibitem[{{Arellano Ferro} {et~al.}(2010){Arellano Ferro}, {Giridhar}, \& {Bramich}}]{Arellano2010}
{Arellano Ferro}, A., {Giridhar}, S., \& {Bramich}, D.~M. 2010, \mnras, 402, 226


\bibitem[{{Arellano Ferro} {et~al.}(2016){Arellano Ferro}, {Luna}, {Bramich}, {Giridhar}, {Ahumada}, \& {Muneer}}]{Arellano2016}
{Arellano Ferro}, A., {Luna}, A., {Bramich}, D.~M., {Giridhar}, S., {Ahumada}, J.~A., \& {Muneer}, S. 2016, \apss, 361, 175


\bibitem[{{Arellano Ferro} {et~al.}(2015){Arellano Ferro}, {Mancera Pi{\~n}a}, {Bramich}, {Giridhar}, {Ahumada}, {Kains}, \& {Kuppuswamy}}]{Arellano2015}
{Arellano Ferro}, A., {Mancera Pi{\~n}a}, P.~E., {Bramich}, D.~M., {Giridhar}, S., {Ahumada}, J.~A., {Kains}, N., \& {Kuppuswamy}, K. 2015, \mnras, 452, 727


\bibitem[{{Arellano Ferro} {et~al.}(2024){Arellano Ferro}, {Zerpa Guillen}, {Yepez}, {Bustos Fierro}, {Prudil}, \& {P{\'e}rez Parra}}]{Arellano2024b}
{Arellano Ferro}, A., {Zerpa Guillen}, L.~J., {Yepez}, M.~A., {Bustos Fierro}, I.~H., {Prudil}, Z., \& {P{\'e}rez Parra}, C.~E. 2024, arXiv e-prints, arXiv:2407.00523


\bibitem[{{Belokurov} {et~al.}(2018){Belokurov}, {Erkal}, {Evans}, {Koposov}, \& {Deason}}]{Belokurov2018}
{Belokurov}, V., {Erkal}, D., {Evans}, N.~W., {Koposov}, S.~E., \& {Deason}, A.~J. 2018, \mnras, 478, 611


\bibitem[{{Bono} {et~al.}(1994){Bono}, {Caputo}, \& {Stellingwerf}}]{Bono1994}
{Bono}, G., {Caputo}, F., \& {Stellingwerf}, R.~F. 1994, \apj, 423, 294


\bibitem[{{Bramich}(2008)}]{Bramich2008}
{Bramich}, D.~M. 2008, \mnras, 386, L77


\bibitem[{{Bramich} {et~al.}(2015){Bramich}, {Bachelet}, {Alsubai}, {Mislis}, \& {Parley}}]{Bramich2015}
{Bramich}, D.~M., {Bachelet}, E., {Alsubai}, K.~A., {Mislis}, D., \& {Parley}, N. 2015, \aap, 577, A108


\bibitem[{{Bramich} {et~al.}(2013){Bramich}, {Horne}, {Albrow}, {Tsapras}, {Snodgrass}, {Street}, {Hundertmark}, {Kains}, {Arellano Ferro}, {Figuera}, \& {Giridhar}}]{Bramich2013}
{Bramich}, D.~M., {Horne}, K., {Albrow}, M.~D., {Tsapras}, Y., {Snodgrass}, C., {Street}, R.~A., {Hundertmark}, M., {Kains}, N., {Arellano Ferro}, A., {Figuera}, J.~R., \& {Giridhar}, S. 2013, \mnras, 428, 2275


\bibitem[{{Bustos Fierro} \& {Calder{\'o}n}(2019)}]{Bustos2019}
{Bustos Fierro}, I.~H. \& {Calder{\'o}n}, J.~H. 2019, \mnras, 488, 3024


\bibitem[{{Callingham} {et~al.}(2022){Callingham}, {Cautun}, {Deason}, {Frenk}, {Grand}, \& {Marinacci}}]{Callingham2022}
{Callingham}, T.~M., {Cautun}, M., {Deason}, A.~J., {Frenk}, C.~S., {Grand}, R. J.~J., \& {Marinacci}, F. 2022, \mnras, 513, 4107


\bibitem[{{Carballo-Bello} {et~al.}(2018){Carballo-Bello}, {Mart{\'\i}nez-Delgado}, {Navarrete}, {Catelan}, {Mu{\~n}oz}, {Antoja}, \& {Sollima}}]{Carballo2018}
{Carballo-Bello}, J.~A., {Mart{\'\i}nez-Delgado}, D., {Navarrete}, C., {Catelan}, M., {Mu{\~n}oz}, R.~R., {Antoja}, T., \& {Sollima}, A. 2018, \mnras, 474, 683


\bibitem[{{Carretta} {et~al.}(2009){Carretta}, {Bragaglia}, {Gratton}, {D'Orazi}, \& {Lucatello}}]{Carretta2009}
{Carretta}, E., {Bragaglia}, A., {Gratton}, R., {D'Orazi}, V., \& {Lucatello}, S. 2009, \aap, 508, 695


\bibitem[{{Catelan}(2009)}]{Catelan2009}
{Catelan}, M. 2009, \apss, 320, 261


\bibitem[{{Clement} {et~al.}(2001){Clement}, {Muzzin}, {Dufton}, {Ponnampalam}, {Wang}, {Burford}, {Richardson}, {Rosebery}, {Rowe}, \& {Hogg}}]{Clement2001}
{Clement}, C.~M., {Muzzin}, A., {Dufton}, Q., {Ponnampalam}, T., {Wang}, J., {Burford}, J., {Richardson}, A., {Rosebery}, T., {Rowe}, J., \& {Hogg}, H.~S. 2001, \aj, 122, 2587


\bibitem[{{Demarque} {et~al.}(2000){Demarque}, {Zinn}, {Lee}, \& {Yi}}]{Demarque2000}
{Demarque}, P., {Zinn}, R., {Lee}, Y.-W., \& {Yi}, S. 2000, \aj, 119, 1398


\bibitem[{{Gaia Collaboration} {et~al.}(2023){Gaia Collaboration}, {Vallenari}, {Brown}, {Prusti}, \& et~al.}]{Gaia2023}
{Gaia Collaboration}, {Vallenari}, A., {Brown}, A.~G.~A., {Prusti}, T., \& et~al. 2023, \aap, 674, A1


\bibitem[{{Guldenschuh} {et~al.}(2005){Guldenschuh}, {Layden}, {Wan}, {Whiting}, {van der Bliek}, {Baca}, {Carlin}, {Freismuth}, {Mora}, {Salyk}, {Vera}, {Verdugo}, \& {Young}}]{Guldenschuh2005}
{Guldenschuh}, K.~A., {Layden}, A.~C., {Wan}, Y., {Whiting}, A., {van der Bliek}, N., {Baca}, P., {Carlin}, J., {Freismuth}, T., {Mora}, M., {Salyk}, C., {Vera}, S., {Verdugo}, M., \& {Young}, A. 2005, \pasp, 117, 721


\bibitem[{{Harris}(1996)}]{Harris1996}
{Harris}, W.~E. 1996, \aj, 112, 1487


\bibitem[{{Helmi} {et~al.}(2018){Helmi}, {Babusiaux}, {Koppelman}, {Massari}, {Veljanoski}, \& {Brown}}]{Helmi2018}
{Helmi}, A., {Babusiaux}, C., {Koppelman}, H.~H., {Massari}, D., {Veljanoski}, J., \& {Brown}, A. G.~A. 2018, \nat, 563, 85


\bibitem[{{Koleva} {et~al.}(2008){Koleva}, {Prugniel}, {Ocvirk}, {Le Borgne}, \& {Soubiran}}]{Koleva2008}
{Koleva}, M., {Prugniel}, P., {Ocvirk}, P., {Le Borgne}, D., \& {Soubiran}, C. 2008, \mnras, 385, 1998


\bibitem[{{Kuzma} {et~al.}(2018){Kuzma}, {Da Costa}, \& {Mackey}}]{Kuzma2018}
{Kuzma}, P.~B., {Da Costa}, G.~S., \& {Mackey}, A.~D. 2018, \mnras, 473, 2881


\bibitem[{{Landolt}(1992)}]{Landolt1992}
{Landolt}, A.~U. 1992, \aj, 104, 340


\bibitem[{{Layden} {et~al.}(2010){Layden}, {Broderick}, {Pohl}, {Reichart}, {Ivarsen}, {Haislip}, {Nysewander}, {LaCluyze}, \& {Corwin}}]{Layden2010}
{Layden}, A.~C., {Broderick}, A.~J., {Pohl}, B.~L., {Reichart}, D.~E., {Ivarsen}, K.~M., {Haislip}, J.~B., {Nysewander}, M.~C., {LaCluyze}, A.~P., \& {Corwin}, T.~M. 2010, \pasp, 122, 1000


\bibitem[{{Lee} {et~al.}(1994){Lee}, {Demarque}, \& {Zinn}}]{Lee1994}
{Lee}, Y.-W., {Demarque}, P., \& {Zinn}, R. 1994, \apj, 423, 248


\bibitem[{{Marino} {et~al.}(2014){Marino}, {Milone}, {Yong}, {Dotter}, {Da Costa}, {Asplund}, {Jerjen}, {Mackey}, {Norris}, {Cassisi}, {Sbordone}, {Stetson}, {Weiss}, {Aparicio}, {Bedin}, {Lind}, {Monelli}, {Piotto}, {Angeloni}, \& {Buonanno}}]{Marino2014}
{Marino}, A.~F., {Milone}, A.~P., {Yong}, D., {Dotter}, A., {Da Costa}, G., {Asplund}, M., {Jerjen}, H., {Mackey}, D., {Norris}, J., {Cassisi}, S., {Sbordone}, L., {Stetson}, P.~B., {Weiss}, A., {Aparicio}, A., {Bedin}, L.~R., {Lind}, K., {Monelli}, M., {Piotto}, G., {Angeloni}, R., \& {Buonanno}, R. 2014, \mnras, 442, 3044


\bibitem[{{Martin} {et~al.}(2004){Martin}, {Ibata}, {Bellazzini}, {Irwin}, {Lewis}, \& {Dehnen}}]{MArtin2004}
{Martin}, N.~F., {Ibata}, R.~A., {Bellazzini}, M., {Irwin}, M.~J., {Lewis}, G.~F., \& {Dehnen}, W. 2004, \mnras, 348, 12


\bibitem[{{Pols} {et~al.}(1998){Pols}, {Schr{\"o}der}, {Hurley}, {Tout}, \& {Eggleton}}]{Pols1998}
{Pols}, O.~R., {Schr{\"o}der}, K.-P., {Hurley}, J.~R., {Tout}, C.~A., \& {Eggleton}, P.~P. 1998, MNRAS, 298, 525


\bibitem[{{Pols} {et~al.}(1997){Pols}, {Tout}, {Schroder}, {Eggleton}, \& {Manners}}]{Pols1997}
{Pols}, O.~R., {Tout}, C.~A., {Schroder}, K.-P., {Eggleton}, P.~P., \& {Manners}, J. 1997, MNRAS, 289, 869


\bibitem[{{Riello} {et~al.}(2021){Riello}, {De Angeli}, {Evans}, \& {et al.}}]{Riello2021}
{Riello}, M., {De Angeli}, F., {Evans}, D.~W., \& {et al.} 2021, \aap, 649, A3


\bibitem[{{Sawyer}(1939)}]{Sawyer1939}
{Sawyer}, H.~B. 1939, Publications of the David Dunlap Observatory, 1, 125


\bibitem[{{Schlafly} \& {Finkbeiner}(2011)}]{Schlafly2011}
{Schlafly}, E.~F. \& {Finkbeiner}, D.~P. 2011, \apj, 737, 103


\bibitem[{{Schlegel} {et~al.}(1998){Schlegel}, {Finkbeiner}, \& {Davis}}]{Schlegel1998}
{Schlegel}, D.~J., {Finkbeiner}, D.~P., \& {Davis}, M. 1998, \apj, 500, 525


\bibitem[{{Schr{\"o}der} {et~al.}(1997){Schr{\"o}der}, {Pols}, \& {Eggleton}}]{KPS1997}
{Schr{\"o}der}, K.-P., {Pols}, O.~R., \& {Eggleton}, P.~P. 1997, MNRAS, 285, 696


\bibitem[{{Stetson}(2000)}]{Stetson2000}
{Stetson}, P.~B. 2000, \pasp, 112, 925


\bibitem[{{Sturch}(1966)}]{Sturch1966}
{Sturch}, C. 1966, \apj, 143, 774


\bibitem[{{Sumerel} {et~al.}(2004){Sumerel}, {Corwin}, {Catelan}, {Borissova}, \& {Smith}}]{Sumerel2004}
{Sumerel}, A.~N., {Corwin}, T.~M., {Catelan}, M., {Borissova}, J., \& {Smith}, H.~A. 2004, Information Bulletin on Variable Stars, 5533, 1


\bibitem[{{Torelli} {et~al.}(2019){Torelli}, {Iannicola}, {Stetson}, {Ferraro}, {Bono}, {Salaris}, {Castellani}, {Dall'Ora}, {Fontana}, {Monelli}, \& {Pietrinferni}}]{Torelli2019}
{Torelli}, M., {Iannicola}, G., {Stetson}, P.~B., {Ferraro}, I., {Bono}, G., {Salaris}, M., {Castellani}, M., {Dall'Ora}, M., {Fontana}, A., {Monelli}, M., \& {Pietrinferni}, A. 2019, \aap, 629, A53


\bibitem[{{VandenBerg} {et~al.}(2013){VandenBerg}, {Brogaard}, {Leaman}, \& {Casagrande}}]{VandenBerg2013}
{VandenBerg}, D.~A., {Brogaard}, K., {Leaman}, R., \& {Casagrande}, L. 2013, \apj, 775, 134


\bibitem[{{Vasiliev} \& {Baumgardt}(2021)}]{Vasiliev2021}
{Vasiliev}, E. \& {Baumgardt}, H. 2021, \mnras, 505, 5978


\bibitem[{{Walker}(1998)}]{WALKER1998}
{Walker}, A.~R. 1998, \aj, 116, 220


\bibitem[{{Yepez} {et~al.}(2022){Yepez}, {Arellano Ferro}, {Deras}, {Bustos Fierro}, {Muneer}, \& {Schr{\"o}der}}]{Yepez2022}
{Yepez}, M.~A., {Arellano Ferro}, A., {Deras}, D., {Bustos Fierro}, I., {Muneer}, S., \& {Schr{\"o}der}, K.~P. 2022, \mnras


\bibitem[{Zhang {et~al.}(1996)Zhang, Ramakrishnan, \& Livny}]{Zhang1996}
Zhang, T., Ramakrishnan, R., \& Livny, M. 1996, SIGMOD Rec., 25, 103
 \href{http://doi.acm.org/10.1145/235968.233324}{\urllinklabel}

\bibitem[{{Zinn} \& {West}(1984)}]{Zinn1984}
{Zinn}, R. \& {West}, M.~J. 1984, \apjs, 55, 45


\end{thebibliography}

\appendix
\section{APPENDIX}
\setcounter{figure}{0}
\counterwithin{figure}{section}

\subsection{Light curves of measured variable stars}

\begin{figure*}[ht]
\begin{center}
\includegraphics[width=16.0cm]{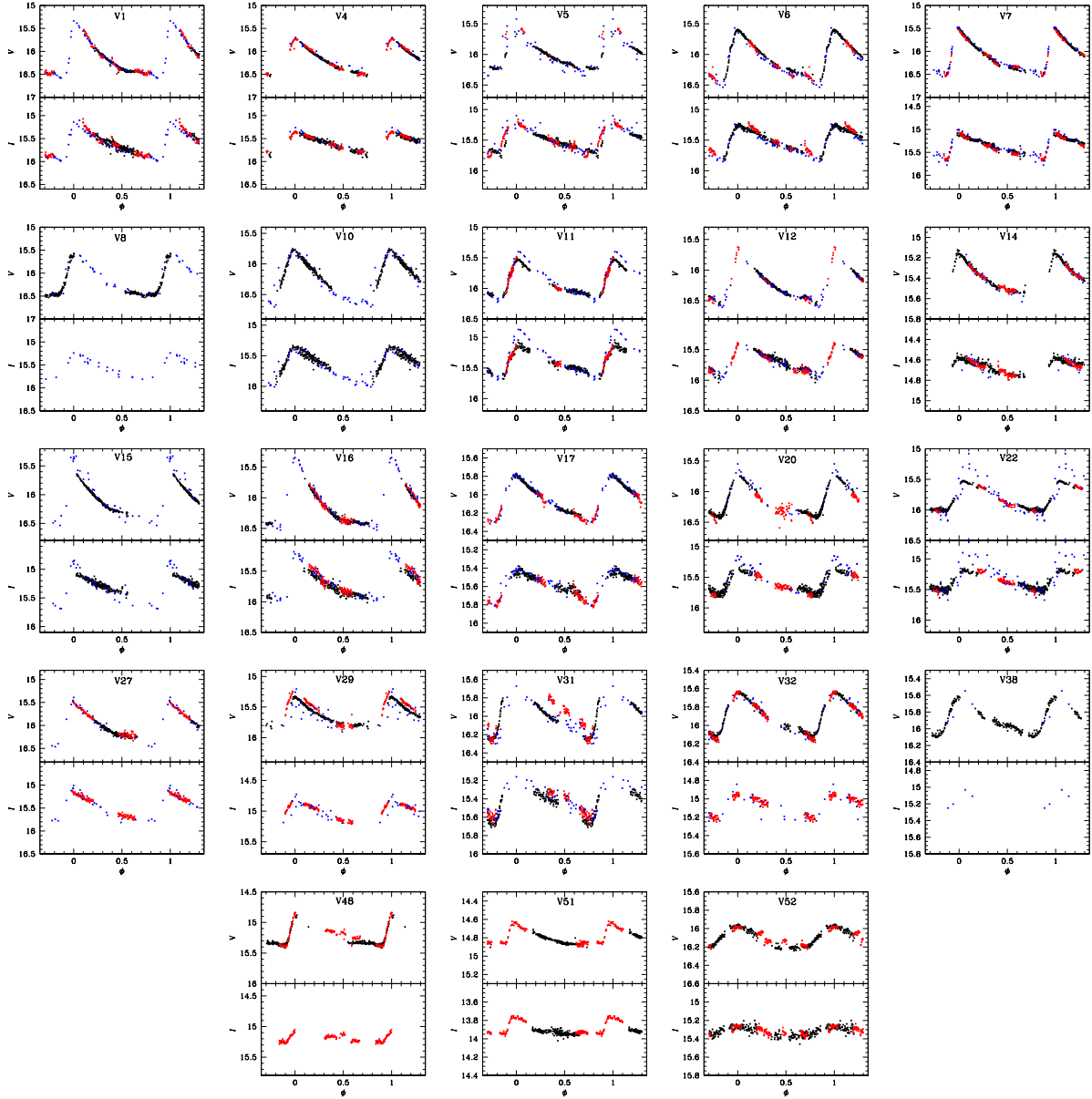}
\caption{NGC 1851 RR Lyrae stars \emph{VI} light curves. The colour code is:, black: BA18, red; BA19, blue: $Gaia$-DR3.}
\label{Mosai_RRab}
\end{center}
\end{figure*}

\begin{figure*}[htp]
\begin{center}
\includegraphics[width=16.0cm]{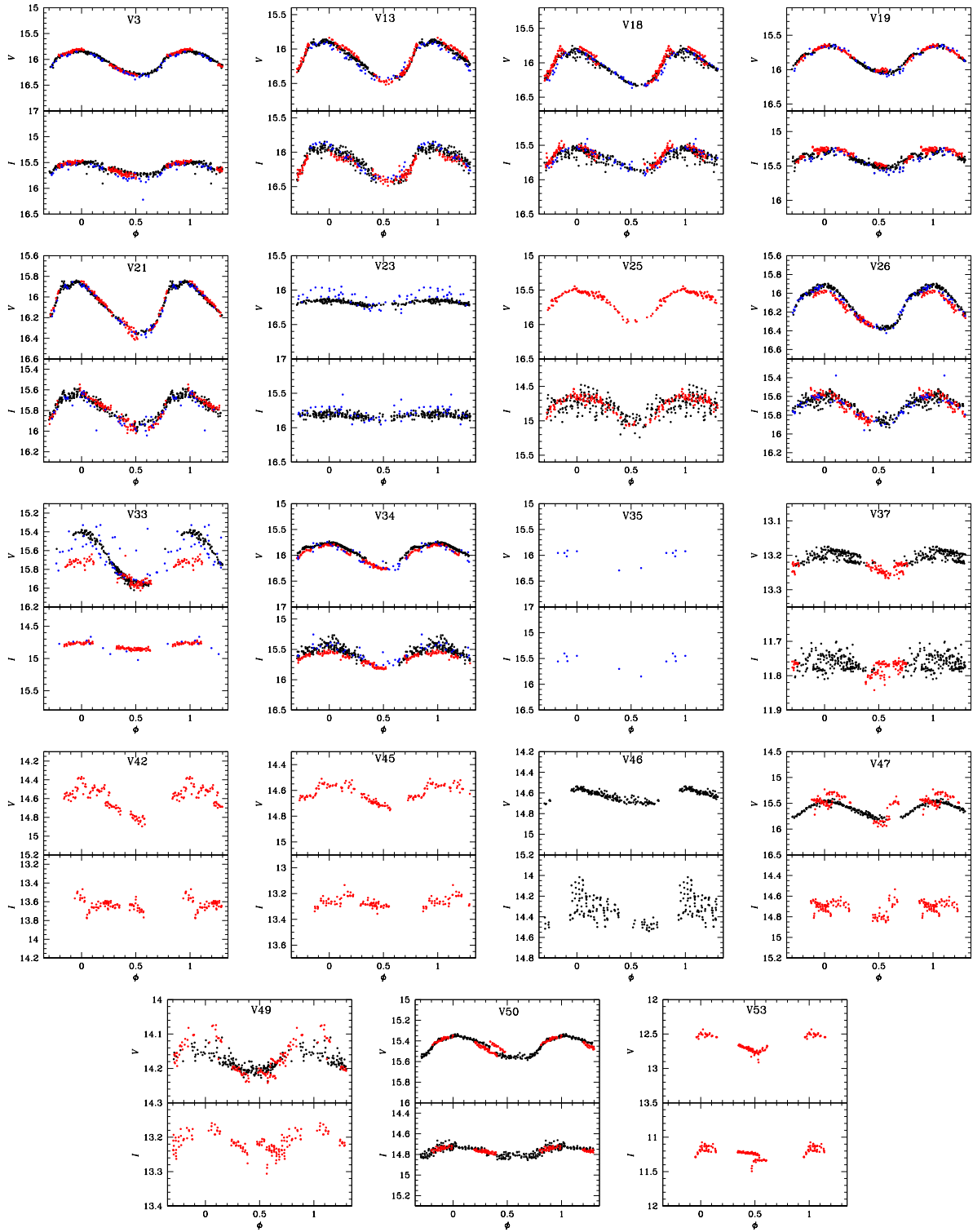}
\caption{Light curves of the RRc stars in th efield of NGC 1851.
The colour code is as Fig. \ref{Mosai_RRab}}
\label{Mosai_RRc}
\end{center}
\end{figure*}

\begin{figure*}[ht]
\begin{center}
\includegraphics[width=16.0cm]{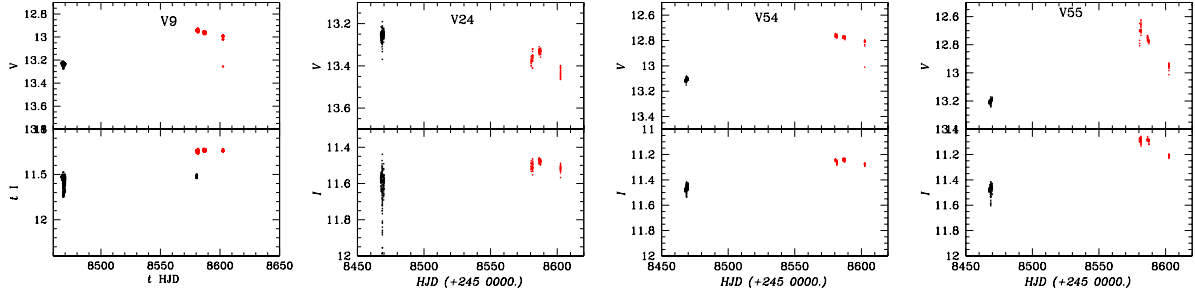}
\caption{Light curves of long-period variables in NGC 1851 phased with the periods from Table\ref{tab:datosgenerales1}. Color coding follows that of . \ref{Mosai_RRab}}
\label{Mosai_SR}
\end{center}
\end{figure*}

\begin{figure*}[htp]
\begin{center}
\includegraphics[width=16.0cm]{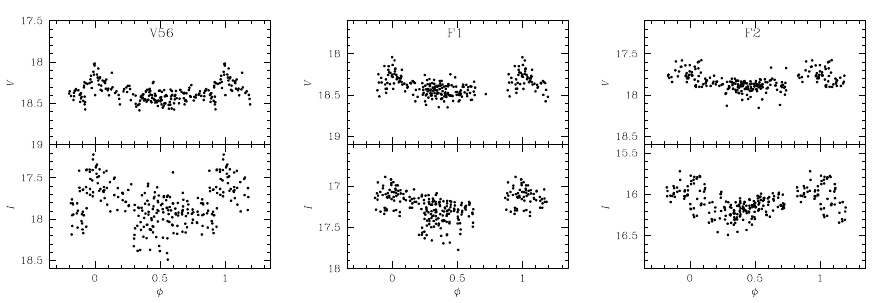}
\caption{We identified new variable stars that had not been previously recorded.} 
\label{MOSAI_Nueva}
\end{center}
\end{figure*}

\begin{figure*}[htp]
\begin{center}
\includegraphics[width=16.0cm]{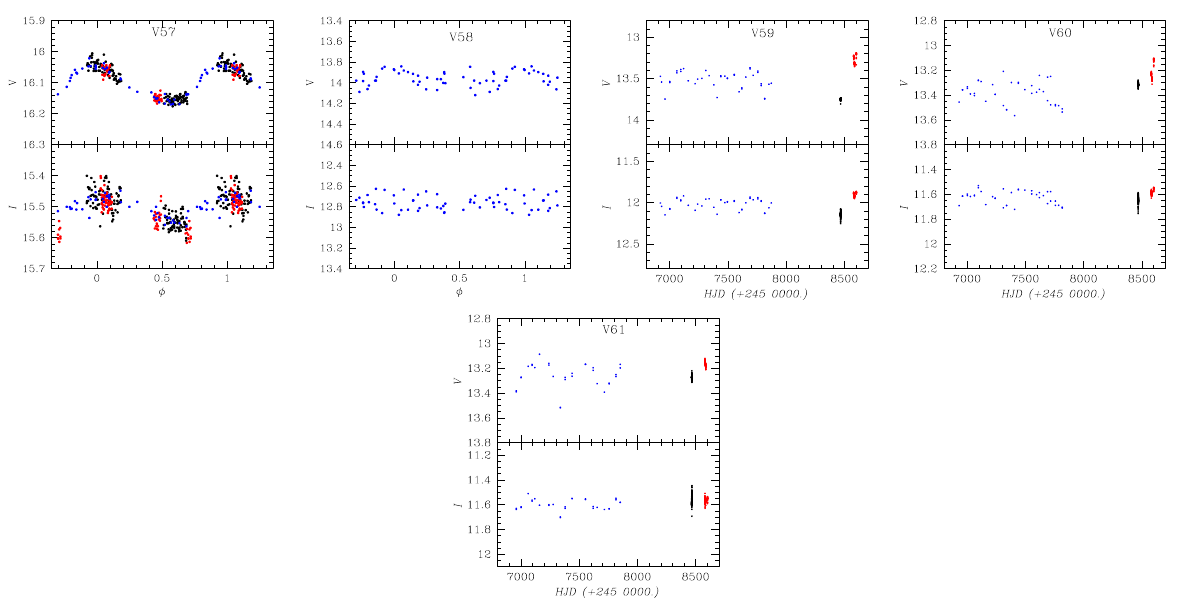}
\caption{Distribution of the confirmed stellar variables by Gaia in this work.}
\label{MOSAI_5gaia}
\end{center}
\end{figure*}

The light curves of all variables resolved in our photometry are displayed in Figs. \ref{Mosai_RRab}, \ref{Mosai_RRc}, \ref{Mosai_SR},\ref{MOSAI_Nueva} and  \ref{MOSAI_5gaia} for the RRab, RRc, RGBs, newly detected variables  and confirmed $Gaia$ variables, respectively.

\subsection{Comments on individual stars}

V13. This RRc star falls too far to the blue of the HB. Its membership to the the cluster is controversial. The assignment by the B\&C membership method and the membership probabilty assigned by V\&B is contradictory. In the Period-amplitud diagram the star has too large an amplitude for its period.

V25. In spite of being an eclipsing binary EC, its light curve is included in Fig. \ref{Mosai_RRc}. The only difference with other RRc stars is that its period is much shorter, 0.173673 d, and it is brighter than the HB by about half a magnitude (see CMD of Fig. \ref{CMD}), The $V$ light curve exhibits a small flattening near maximum which may be an incipient suggestion of a secondary eclipse. This is also seen in the $I$-band light curve from the BA19 season. 

V34. This variable was first reported by \citet{Sumerel2004} as an RRab star with a period of 0.515 d, which in fact produces a well phased light curve.  However, at the coordinates given in the CVSGC we in fact find a variable star, but we find a period of 0.345033 d that displays a clear and complete RRc-like light curve (see Fig.  \ref{Mosai_RRc}). The only reason we find for this discrepancy is that the data of \citet{Sumerel2004} cover only about half a cycle, and then their period and type may be spurious. We classified the star as RRc.

V51. It was reported as variable by \citet{Sumerel2004}, and the light curve reported by these authors (labeled NV18), although incomplete, clearly suggests the RRab nature of the star. It was noticed by \citet{Layden2010} that the star is in fact an RR Lyrae stars badly blended with a non variable stars previously identified as V2 by \citet{Sawyer1939}. The light curve measured by Layden (2010)  was not published but it was said to be noisy, likely due to the contamination of the brighter V2. Our light curve in Fig. \ref{Mosai_RRab} is fairly complete, confirms the RRab nature of the star nicely phased with a period of 0.509389 days. The mean \emph{VI} magnitude level of our curve is spurious due to the light contamination of V2 and it is its position in the CMD of Fig. \ref{CMD}.

V52. This star is not classified in the present edition of the CVSGC where only its X-Y coordinates are listed. We have identified the star and found it to be a cluster member RRab star. It sits on the HB and its light curve is properly phased with a period of 0.648831 d.

V56. This is a newly identified variable in this work. Its light curve shape and position on the CMD diagram remind us of an SX Phe type star. However a period of 0.25 d. is perhaps a little too long for an SX Phe. We have retained it in our general Table 4 as SX Phe? awaiting a confirmation in the future.

V58. Its light curve is that of an RRab, however it appears about two magnitudes above the HB on near the RGB. Both membership approaches, B\&C and V\&B, based on its proper motion, identify the star as a cluster member. In the identification chart we see that evidently the star is blended with at least a brighter star which explains itsmean magnitude to be spuriously too bright. We have considered the star to be a cluster member and assigned the variable name V58.
\end{document}